\documentclass{Interspeech}
\usepackage{makecell}
\usepackage{marvosym}

\interspeechcameraready

\title{The Multimodal Information Based Speech Processing (MISP) 2025 Challenge: Audio-Visual Diarization and  Recognition}

\author[affiliation={1}]{Ming}{Gao}
\author[affiliation={1}]{Shilong}{Wu}
\author[affiliation={1}]{Hang}{Chen}
\author[affiliation={1,*}]{Jun}{Du}
\author[affiliation={2}]{Chin-Hui}{Lee}
\author[affiliation={3}]{Shinji}{Watanabe}
\author[affiliation={4}]{Jingdong}{Chen}
\author[affiliation={5}]{Siniscalchi Sabato}{Marco}
\author[affiliation={6}]{Odette}{Scharenborg}

\affiliation{}{University of Science and Technology of China}{China $^2$Georgia Institute of Technology, USA $^3$Carnegie Mellon University, USA $^4$Northwestern Polytechnical University, China $^5$University of Palermo, Italy $^6$Delft University of Technology, Netherlands}

\email{\Letter jundu@ustc.edu.cn}

\newcommand{\blue}[1]{\textcolor{blue}{#1}}
\usepackage{comment}

\begin{document}

\maketitle

\begin{abstract}
    
    Meetings are a valuable yet challenging scenario for speech applications due to complex acoustic conditions. This paper summarizes the outcomes of the MISP 2025 Challenge, hosted at Interspeech 2025, which focuses on multi-modal, multi-device meeting transcription by incorporating video modality alongside audio. The tasks include Audio-Visual Speaker Diarization (AVSD), Audio-Visual Speech Recognition (AVSR), and Audio-Visual Diarization and Recognition (AVDR). We present the challenge’s objectives, tasks, dataset, baseline systems, and solutions proposed by participants. The best-performing systems achieved significant improvements over the baseline: the top AVSD model achieved a Diarization Error Rate (DER) of 8.09\%, improving by 7.43\%; the top AVSR system achieved a Character Error Rate (CER) of 9.48\%, improving by 10.62\%; and the best AVDR system achieved a concatenated minimum-permutation Character Error Rate (cpCER) of 11.56\%, improving by 72.49\%.
\end{abstract}

\section{Introduction}

In recent years, the proliferation of speech-enabled applications has led to increasingly complex usage scenarios, such as home environments, and professional meetings. These scenarios present considerable challenges due to adverse acoustic conditions, including far-field audio, background noise, and reverberation. Additionally, conversational interactions often involve multiple speakers with substantial speech overlap. Most publicly available meeting corpora have limited scope and lack summary annotations. For instance, the CHIL dataset\cite{chil} contains only 20 English meetings with 80 speakers and 72 hours of content. Audio-only datasets like AliMeeting\cite{ali} and Aishell-4\cite{aishell4} include 500 and 60 Mandarin meetings, respectively. The simulated LibriCSS dataset\cite{css} recreates meeting dynamics by playing LibriSpeech\cite{librispeech} utterances through loudspeakers, but the dialogues lack the continuity of real conversations.

Current audio-only speech processing techniques are facing performance plateaus, with systems like those in CHiME-6\cite{watanabe20b_chime} achieving a word error rate (WER) of 30\%, and a character error rate (CER) of 20\%, limiting their real-world deployment. The McGurk effect and subsequent studies show that visual cues, such as lip movements, enhance speech perception, particularly in noisy environments. Motivated by this, the MISP challenge aims to advance speech processing by incorporating additional modalities, namely video.

Previous MISP 2021\cite{misp2021}, 2022\cite{misp2022} and 2023\cite{misp2023} challenges targeted the home scenario, where several people chatted in Chinese while watching TV in a living room. A large-scale audio-visual Chinese home conversational corpus was released to support tasks like wake word spotting, target speaker extraction, speaker diarization, and speech recognition. It is the first and largest distant multi-microphone Chinese audio-visual corpus. The dataset’s availability led to over 150 teams downloading it, more than 60 teams submitting results, and 15 papers presented at ICASSP 2022-2024. Motivated by the success of previous MISP  challenges, the MISP 2025 challenge focuses on multi-modal multi-device meeting transcription and aims to push the boundaries of current techniques by introducing additional modality information, namely the video modality. The specific tasks considered in the challenge are Audio-Visual Speaker Diarization (AVSD), Audio-Visual Speech Recognition (AVSR) and Audio-Visual Diarization and Recognition (AVDR).

\begin{table}[!t]
\centering
\caption{Details of MISP-Meeting corpus}
\label{tab:total_statistics}
\setlength{\tabcolsep}{6pt}
\begin{tabular}{lcccc}
\toprule
\textbf{Dataset}& \textbf{Train} & \textbf{Dev} & \textbf{Eval} & \textbf{Total}\\
\midrule
\textbf{Duration (h)} & $119$ & $3$ & $3$ & $125$ \\
\textbf{Session} & $72$ & $9$ & $9$ & $90$ \\
\textbf{Room} & $15$ & $4$ & $4$ & $23$ \\
\textbf{Participant} & $233$ & $15$ & $15$ & $263$ \\
\textbf{\quad -Male} & $115$ & $7$ & $8$ & $130$ \\
\textbf{\quad -Female} & $118$ & $8$ & $7$ & $133$ \\
\textbf{Overlap Ratio} & $57.30\%$ & $46.54\%$ & $53.61\%$ & $56.95\%$ \\
\bottomrule
\end{tabular}
\end{table}

This paper makes the following key contributions to the field of multimodal speech processing:
\begin{itemize}
\item \textbf{An innovative audio-visual speech recognition challenge (MISP 2025).} We propose a new challenge focused on audio-visual speech recognition, providing a large-scale, multi-device, multi-modal dataset specifically designed for meeting scenarios. The challenge includes synchronized audio and video recordings captured under realistic conditions, addressing the need for high-quality resources in this domain.

\item \textbf{A comprehensive baseline system.} We provide a complete set of baseline implementations for the tasks of audio-visual speaker diarization, audio-visual speech recognition, and joint diarization and recognition. These baselines are designed to facilitate rapid experimentation and benchmarking by researchers and practitioners.

\item \textbf{In-depth analysis of participant methods.} We present a thorough summary and analysis of the techniques submitted by participants in the MISP 2025 Challenge. That includes a discussion of effective approaches, their strengths and limitations, and insights into promising directions for future.
\end{itemize}

%By providing these resources and analyses, this paper aims to serve as a foundational reference for the research community, enabling further advancements in multimodal speech processing for complex real-world scenarios.

\begin{comment}
\begin{table}[!t]
\centering
\caption{Statistics of meeting rooms}
\label{tab:room}
\setlength{\tabcolsep}{5pt}
\begin{tabular}{lccccc}
\toprule
\textbf{Size}& \textbf{Tiny}  & \textbf{Small} & \textbf{Middle} & \textbf{Large} & \textbf{Total}\\
\textbf{Area} (in $\rm{m}^2$)& $0$-$18$ & $18$-$36$ & $36$-$60$ & $60$-$\infty$ & $0$-$\infty$\\
\midrule
\textbf{Train} & $5$ & $5$ & $3$ & $2$ & $15$\\
\textbf{Dev} & $1$ & $1$ & $1$ & $1$ & $4$\\
\textbf{Eval} & $1$ & $1$ & $1$ & $1$ & $4$\\
\bottomrule
\end{tabular}
\end{table}
\end{comment}

\begin{figure}[!t]
\centering
%\vspace{-10mm}
\includegraphics[width=1.0\columnwidth]{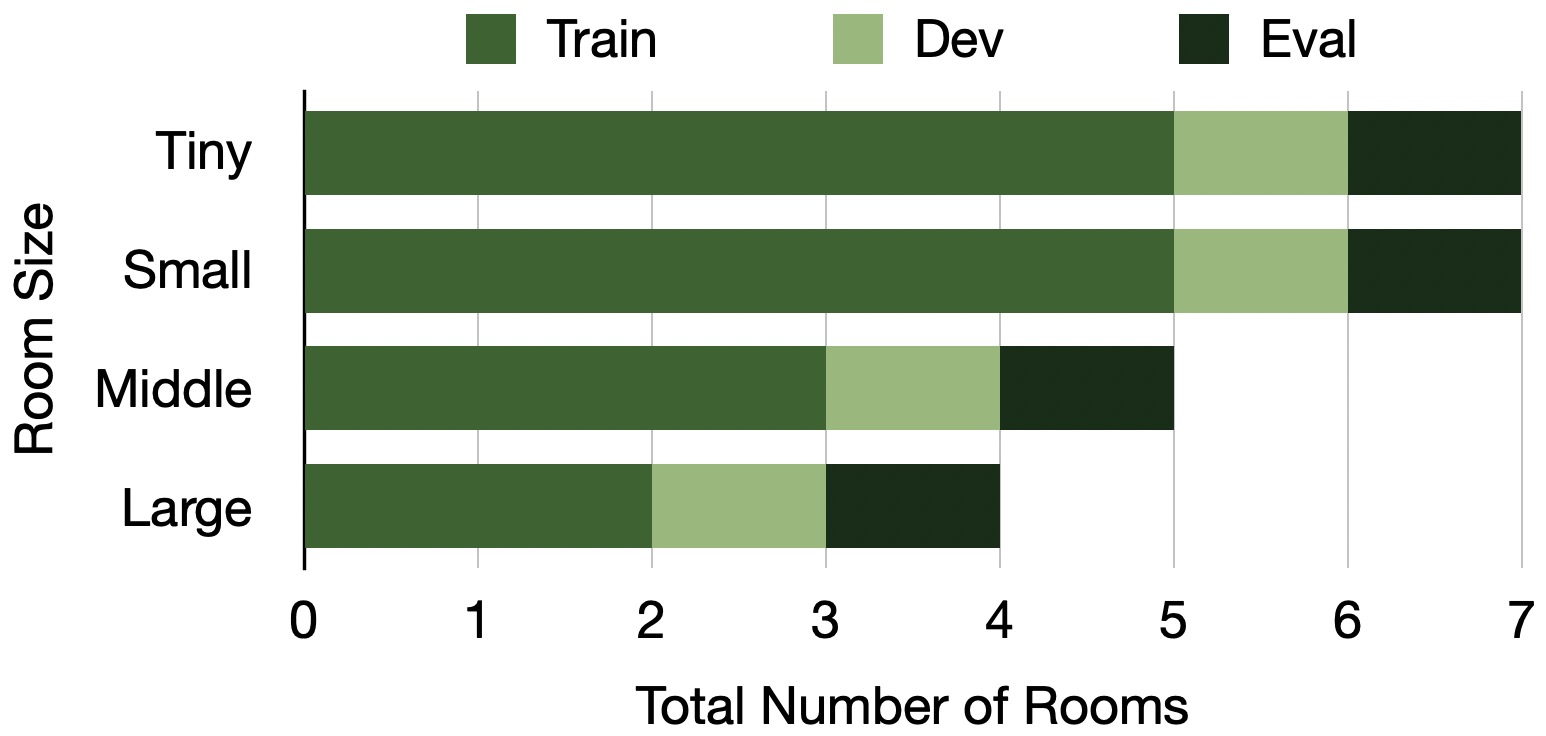}
\caption{Statistics of meeting rooms.}
\label{fig:meetingroom}
\end{figure}

\section{Dataset}
\subsection{Statics}
The MISP-Meeting dataset\cite{misp2025} comprises a total of 125 hours of synchronized audio and video data, meticulously curated to reflect real-world meeting scenarios. The dataset is partitioned into three distinct subsets: 119 hours for training (Train), 3 hours for development (Dev), and 3 hours for evaluation (Eval), the latter serving as the basis for rigorous scoring and ranking. Specifically, the training, development, and evaluation sets consist of 72, 9, and 9 sessions, respectively. To ensure robustness and generalizability, there is no overlap in speakers or recording rooms across these subsets. Each session involves 4-8 participants engaged in natural discussions, with session durations tailored to the subset: training sessions span 2 hours, while development and evaluation sessions are condensed to 20 minutes. This design ensures that training sessions encompass multiple topic transitions, simulating the dynamic nature of real meetings. The dataset includes a total of 233 participants in the training set, and 15 participants each in the development and evaluation sets, with a balanced gender representation. Notably, all participants are professionals, or students whose fields of expertise align closely with the meeting topics, enhancing the authenticity of the discussions while minimizing prolonged silent intervals. The proportion of overlapping speech segments in the training, development, and evaluation sets is 57.30\%, reflecting the natural conversational dynamics of real-world meetings.Details of the dataset, including session counts, participant demographics, and speech overlap statistics, is in Table~\ref{tab:total_statistics}.

A distinguishing feature of the MISP-Meeting corpus is the diversity of its meeting room environments. As summarized in Figure~\ref{fig:meetingroom}, the dataset encompasses 23 meeting rooms categorized into four size groups: tiny (8.79–20 m²), small (20–40 m²), medium (40–80 m²), and large (80–117.6 m²). Each subset includes rooms from all size categories, ensuring a wide range of acoustic properties and spatial configurations. The rooms are further characterized by varied wall materials (e.g., cement, glass) and furnishings (e.g., sofas, TVs, blackboards, fans, air conditioners, plants), which collectively contribute to the acoustic diversity of the dataset. Detailed acoustic parameters for each meeting venue will be released alongside the training data, providing researchers with a valuable resource for in-depth acoustic analysis and model development.

\subsection{Data Collection}

\begin{figure}[!t]
\centering
%\vspace{-10mm}
\includegraphics[width=1.0\columnwidth]{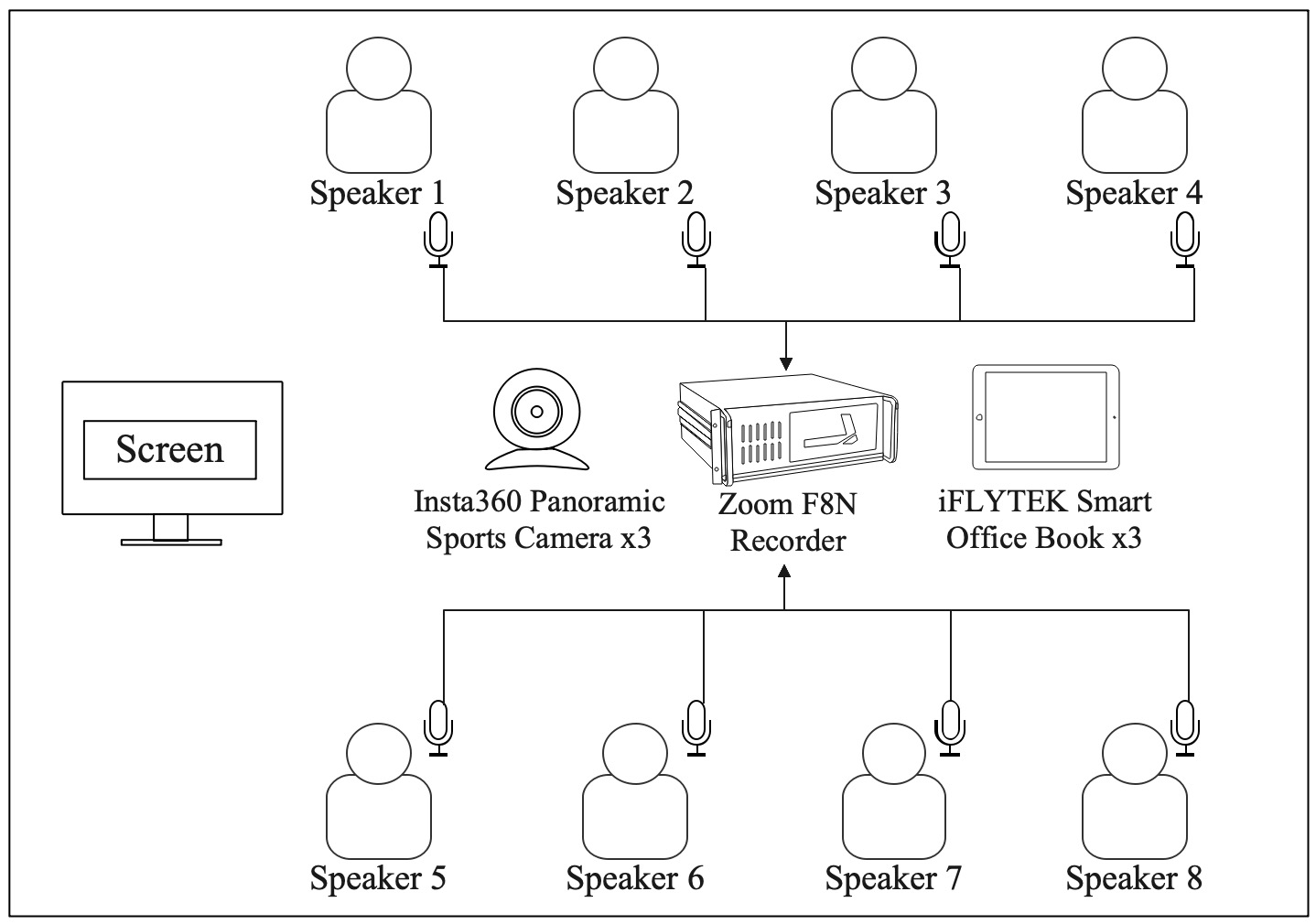}
\caption{Example of recording venue, and used devices.}
\label{fig:meeting}
\end{figure}

As shown in Figure~\ref{fig:meeting}, meeting attendees are seated around an 8-microphone array and a panoramic camera, both placed on the table in a standard meeting room. Participants engage in natural conversations on various topics, including medical treatment, education, business, and industrial production. During the sessions, different indoor noises, such as clicking, keyboard typing, door opening and closing, and fan sounds, naturally occur.

The microphone array is integrated into the iFLYTEK Smart Office Book X3, configured in a rectangular shape with dimensions of 197 mm by 134 mm. It consists of eight omnidirectional microphones symmetrically distributed along the width edges, each capturing audio at a 16 kHz sampling rate and 32-bit resolution. Adjacent to the array is an Insta360 Panoramic Sports Camera X3, measuring 46 mm by 114 mm, with two fisheye lenses on the front and back surfaces and two omnidirectional microphones on the sides. The setup is mounted 30-40 cm above the table surface. The outputs include 360-degree panoramic video in MP4 format, with a resolution of 3840x1920 at 30 fps, and 2-channel audio recorded at 48 kHz and 16-bit resolution.

Each participant wore a headset microphone that captured near-field speech at 44.1 kHz and 16-bit resolution, minimizing interference from off-target sources and ensuring a signal-to-noise ratio (SNR) greater than 15 dB for high-quality manual transcription. All microphones were connected to a Zoom F8N Recorder with a shared clock for synchronized recording. Despite the synchronized microphones, there were still three distinct clocks within the system: the microphone array, the camera, and the recorder. These clocks were manually synchronized by identifying a common reference point, such as knocking a cup. The visual frame capturing the moment of impact, and the corresponding sound waveform were manually aligned using the provided timestamps.

\section{Challenge Description}
%The MISP 2025 Challenge focuses on advancing multimodal speech processing techniques in meeting scenarios by introducing three core tasks: Audio-Visual Speaker Diarization (AVSD), Audio-Visual Speech Recognition (AVSR), and Audio-Visual Diarization and Recognition (AVDR). Each task is designed to address specific challenges in real-world meeting environments, leveraging both audio and visual modalities to improve performance. Below, we describe the tasks, evaluation metrics, and baseline systems in detail.
\subsection{Task 1: Audio-Visual Speaker Diarization (AVSD)}
%\subsubsection{Task 1: Audio-visual speaker diarization}

Audio-visual speaker diarization aims to solve the ``who spoke when'' problem by labeling speech timestamps with classes corresponding to speaker identity using multi-speaker audio and video data. Training and development sets provide all audio and video recordings along with the corresponding ground-truth segmentation timestamp. In contrast, the evaluation set does not include the near-field speech, and the transcriptions. Participants need to determine the speaker at each time point. 

We follow our previous work~\cite{he2022end} to deploy the baseline system, with adaptations to accommodate the far-field audio and video data used in this challenge. The baseline system consists of a visual encoder, audio encoder, and decoder. The visual encoder processes lip region inputs with a lipreading model\cite{26}, Conformer blocks\cite{26}, and a bidirectional LSTM (BLSTM) layer to generate visual embeddings and an initial diarization result. The audio encoder processes the dereverberated audio, using the diarization results from the V-VAD model to compute i-vectors as speaker embeddings, while also generating audio embeddings through CNN layers. The decoder combines those embeddings and uses BLSTM with projection layers to output speech/non-speech probabilities. Furthermore, DOVER-Lap~\cite{29} is used to fuse the results of 8-channels audio. Training occurs in three stages: pre-training the visual network for V-VAD, training the audio network and decoder, and joint fine-tuning of the entire network.

Diarization error rate (DER)~\cite{fiscus2006rich} is adopted as the evaluation criterion. The lower the DER value (with 0 being a perfect score), the higher the ranking. DER is calculated as: the summed time of three different errors of speaker confusion (SC), false alarm (FA), and missed detection (MD) divided by the total duration time, as shown in
\begin{equation}
{\rm DER} = \frac{T_{\rm SC}+T_{\rm FA}+T_{\rm MD}}{T_{\rm total}}
\label{eq:der}
\end{equation}
where $T_{\rm SC}$, $T_{\rm FA}$ and $T_{\rm MD}$ are the time duration of the three errors, and $T_{\rm total}$ is the total time duration. It is worth noting that we do not set the ``no score'' collar, and overlapping speech will be evaluated. 

\subsection{Task 2: Audio-Visual Speech Recognition (AVSR)}
The challenge of AVSR is to handle overlapped segmentation and recognize the content of multiple speakers. The provided evaluation set is the same as Task 1, and the ground truth diarization information is available. Participants are required to transcribe each speaker. 
As for the baseline system, we adopt that from our previous work~\cite{dai2023mproving}. The module uses a hybrid DNN-HMM architecture, combining audio and visual modalities for speech recognition. It starts with preprocessing, where far-field 8-channel audio is dereverberated using GSS~\cite{gss}, and lip ROIs are extracted from video based on diarization results. Lip movement is processed with 3D convolutions and ResNet-18 to generate visual embeddings. Log Mel-filterbank features are passed through 1D convolutions and ResNet-18 to produce audio embeddings.  Audio-visual embeddings are extracted using multi-stage temporal convolutional networks. Posterior probabilities are computed with additional MS-TCN modules.

The performance is measured by character error rate (CER). The CER compares, for a given hypothesis output, the total number of characters, including spaces, to the minimum number of insertions (Ins), substitutions (Subs) and deletions (Del) of characters that are required to obtain the reference transcript. Specifically, CER is calculated by:
\begin{equation}
{\rm CER} = \frac{N_{\rm Subs} + N_{\rm Del} + N_{\rm Ins}}{N_{\rm total}} \times 100
\label{eq:CER}
\end{equation}
where $N_{\rm Subs}$, $N_{\rm Del}$ and $N_{\rm Ins}$ are the character number of the three errors, respectively, and $N_{\rm total}$ is the total number of characters. The lower the CER value (with 0 being a perfect score), the better the recognition performance. For such speech overlap segments, we calculate all errors based on the recognition results and the ground truth for each speaker based on the oracle speaker diarization results.
%Due to the multi-speaker interaction in our scenario, there are speech segments with multiple speakers talking simultaneously. 

\subsection{Task 3: Audio-Visual Diarization and Recognition (AVDR)}

Task 3 aims to directly solve the ``who spoke what when" problem and can be seen as a combination of Task 1 and 2. The evaluation set is the same as Task 2, but using ground truth diarization information is prohibited. 
We cascade the baseline systems from Tasks 1 and 2 to form the baseline system for Task 3, following the same cascading process described in our previous work \cite{wang2023misp}. During inference, the AVSD module outputs an RTTM file with session, speaker, start time (Tstart), and duration (Tdur) information. Each session and speaker’s utterances are segmented based on Tstart and Tdur.  The video frames are cropped to extract lip ROIs, while the audio is dereverberated, beamformed, and segmented.  AVSR  generates transcriptions, which are concatenated in chronological order. 

With reference to the concatenated minimum-permutation word error rate (cpWER) in \cite{watanabe20b_chime}, we use concatenated minimum-permutation character error rate (cpCER) as the evaluation criterion in Task 3. The calculation of cpCER in a session is divided into three steps:
\begin{enumerate}
\item Recognition results and reference transcriptions belonging to the same speaker are concatenated on the timeline in a session.
\item CERs between the reference and all possible speaker permutations of the hypothesis $\{\bm{s}_i | i = 0, 1, \cdots,  \mathrm{P}_{N_{\rm spk}}^{N_{\rm spk}}\}$ are calculated as Eq.~(\ref{eq:CER}), where $N_{\rm spk}$ is the total number of speakers in the session.
\item The lowest CER as the cpCER, the process is as follows:
\begin{equation}
{\rm cpCER} = \min_{\{\bm{s}_i | i = 0, 1, \cdots,  \mathrm{P}_{N_{\rm spk}}^{N_{\rm spk}}\}}{\rm CER}_i
\label{eq:cpCER}
\end{equation}
\end{enumerate}

\section{Results and Analysis}
\subsection{AVSD}

\begin{table*}[!t]
\centering
\caption{Results of submissions of track 1 (Audio-Visual Speech Diarization) in the MISP 2025 challenge}
\label{avsd}
\setlength{\tabcolsep}{5pt}
\begin{tabular}{c|c|cccc|c}
\toprule
\textbf{ID}&\textbf{Team}  & \textbf{Method} & \textbf{Pre-train} & \textbf{External Datasets} &\textbf{Modality} &  \textbf{DER(\%)} \\
\midrule
\textbf{1} & DKU-WHU & MC-SSND & by VoxBlink2\cite{voxblink2} & \thead{VoxCeleb2\cite{voxceleb2}+VoxBlink2\cite{voxblink2}+\\KeSpeech\cite{kespeech}+3D-Speaker\cite{3d}} &  Audio-only &\textbf{8.09}\\
\textbf{2} & XMUSPEECH & CASA-Net & ECAPA-TDNN\cite{tdnn} & MISP 2022\cite{misp2022} & Audio-visual &8.18\\
\textbf{3} & NJU-AALAB & DiariZen & WavLM-large\cite{wavlm} & \thead{AMI\cite{ami}+AISHELL-4\cite{aishell4}+\\AliMeeting\cite{ali}} & Audio-only &8.33\\
\textbf{4} & Fosafer Research & Hybrid-system & WavLM-large\cite{wavlm} & \thead{AISHELL-4\cite{aishell4}+AliMeeting\cite{ali}+\\MISP2021\cite{misp2021}+VoxCeleb2\cite{voxceleb2}} & Audio-only &8.88\\
\textbf{5} & Baseline & MISP-Baseline & LRW\cite{lrw} & - & Audio-visual & 15.52 \\
\bottomrule
\end{tabular}
\end{table*}

\begin{table*}[!t]
\centering
\caption{Results of submissions of track 2 (Audio-Visual Speech Recognition) and track 3 (Audio-Visual Diarization and Recognition) in the MISP 2025 challenge}
\label{avsr}
\setlength{\tabcolsep}{5pt}
\resizebox{1\textwidth}{!}{
\begin{tabular}{c|c|cccc|cc}
\toprule
\textbf{ID}& \textbf{Team}  & \textbf{Method} & \textbf{Pre-train} & \textbf{External Datasets} &\textbf{Modality} & \textbf{CER(\%)}& \textbf{cpCER(\%)}\\
\midrule
\textbf{1} & Fosafer Research & \thead{ASR-Aware\\OA System} & WavLM\cite{wavlm} & \thead{AliMeeting\cite{ali}+KeSpeech\cite{kespeech}+\\MISP2021\cite{misp2021}+VoxCeleb2\cite{voxceleb2}+\\AISHELL-4\cite{aishell4}} &  Audio-only &\textbf{9.48} &\textbf{11.56}\\
\textbf{2} & DKU-WHU & MC-SSND & Whisper\cite{whisper} & - & Audio-only &15.85&16.88\\
\textbf{3} & Baseline & MISP-Baseline & \thead{WenetSpeech\cite{wenetspeech}+\\LRW\cite{lrw}} & - & Audio-visual & 20.10 &84.05\\
\bottomrule
\end{tabular}
}
\end{table*}

Table~\ref{avsd} presents a summary of the methods and results for Track 1 (Audio-Visual Speaker Diarization) in the MISP 2025 Challenge. The table lists the participating teams, their proposed methods, pre-training strategies, external datasets used, the modality employed, and the corresponding Diarization Error Rate (DER) achieved.

The top four teams in Task 1 (Audio-Visual Speaker Diarization) of the MISP 2025 Challenge showcased a range of innovative approaches. The winning team, DKU-WHU, extended the single-channel SSND model\cite{ssnd} to multi-channel audio with a channel attention mechanism, achieving a DER of 8.09\%. The second-place team, XMUSPEECH, leveraged cross-modal integration through their CASA-Net architecture with cross-attention and self-attention mechanisms, resulting in a DER of 8.18\%. NJUAALAB, in third place, combined end-to-end neural diarization with clustering techniques and advanced models like WavLM-Large\cite{wavlm} and Conformer\cite{26}, yielding a DER of 8.33\%. FOSAFER, the fourth-place team, used a hybrid approach combining traditional methods with WavLM-based segmentation, achieving a DER of 8.88\%.

The teams shared a focus on advanced neural architectures such as Conformer and WavLM, along with multi-channel audio processing. DKU-WHU and NJUAALAB emphasized multi-channel fusion, effectively utilizing channel attention and clustering techniques for complex environments.  In contrast, XMUSPEECH and FOSAFER explored multimodal integration with varying success. XMUSPEECH achieved notable improvements through cross-modal attention, while FOSAFER found visual data less impactful when audio alone was sufficient. Post-processing techniques like DOVER-Lap\cite{29} and temporal alignment were common to all teams, highlighting the importance of refining raw predictions.

The comparison suggests several promising directions for future research: first, further exploration of cross-modal attention could enhance audio-visual data integration in noisy or low-quality environments. Second, FOSAFER’s dynamic hybrid approach indicates potential for adaptive systems tailored to varying acoustic conditions, such as different levels of overlapping speech. Third, NJUAALAB’s use of advanced clustering and fusion techniques could improve the aggregation of diarization results, particularly in multi-channel settings. Finally, the limited impact of visual data in FOSAFER’s approach suggests a need for more sophisticated video-based correction methods to complement audio in multimodal systems.

\subsection{AVSR \& AVDR}

Table~\ref{avsr} provides a summary of the participant methods and results for Tracks 2 (Audio-Visual Speech Recognition) and 3 (Audio-Visual Diarization and Recognition) in the MISP 2025 Challenge. The table includes the team names, methods, pre-training strategies, external datasets used, modalities employed, and the corresponding Character Error Rate (CER) and character-phoneme CER (cpCER) values.

Both the FOSAFER and DKU-WHU teams demonstrated state-of-the-art performance in the MISP 2025 Challenge, yet their approaches exhibit distinct methodologies, and shared insights. The FOSAFER team’s ASR-Aware Observation Addition (ASR-Aware OA) system focuses on addressing low SNR challenges by fusing noisy speech, Mossformer2-separated speech, and GSS-separated speech through a weighted fusion mechanism, guided by a sentence-level bridging module leveraging ASR recognition information. This approach, combined with Mossformer2\cite{mossformer2} and Paraformer\cite{paraformer} models, achieved a remarkable CER of 9.48\%, highlighting the effectiveness of audio-centric strategies. Similarly, the DKU-WHU team used a pretrained Whisper model as backbone, integrating WPE\cite{wpe} and GSS\cite{gss} for speech enhancement and separation, achieving a CER of 15.85\% and a cpCER of 16.88\%. Both teams explored visual information but found it insufficient to enhance performance, underscoring the dominance of audio signals in their systems. However, the FOSAFER team’s hybrid approach to speaker diarization, combining traditional methods with a WavLM-based model, contrasts with the DKU-WHU team’s Multi-Channel Sequence-to-Sequence Neural Diarization (MC-SSND) framework, which refines diarization accuracy using multi-channel audio. While both systems excel in handling overlapping speech and speaker identification, the FOSAFER team’s ASR-Aware OA system demonstrates superior performance in speech recognition, likely due to its innovative fusion strategy and ASR-driven optimization. Further exploration of advanced fusion techniques and domain-specific pretraining could enhance robustness in challenging acoustic environments, paving the way for more accurate and adaptable speech processing systems.

\section{Conclusion}
In this paper, we presented the MISP 2025 Challenge, a multimodal speech processing benchmark focused on meeting scenarios. We introduced the large-scale audio-visual meeting dataset (MISP-Meeting), provided comprehensive baseline systems for audio-visual speaker diarization, speech recognition, and joint diarization and recognition, and analyzed participant submissions. Future work will focus on improving multimodal fusion, tackling domain adaptation challenges, and expanding the dataset to cover more diverse meeting scenarios. These efforts will significantly advance multimodal speech processing.

\bibliographystyle{IEEEtran}
\bibliography{mybib}

\end{document}